\documentclass[superscriptaddress,aps,pra,twocolumn,showpacs,nofootinbib,longbibliography, floatfix]{revtex4-1}
\usepackage{amsmath,amsthm}
\usepackage{braket}
\usepackage{amsmath}
\usepackage{amsfonts}
\usepackage{amssymb}\usepackage[T1]{fontenc}
\usepackage{latexsym}
\usepackage{amssymb}
\usepackage[colorlinks=true,citecolor=blue,urlcolor=blue]{hyperref}
\usepackage{color}
\usepackage{graphics,epstopdf}
\usepackage{soul}
\usepackage[demo]{graphicx}
\usepackage{lipsum}
\usepackage{float}
\usepackage{adjustbox}
\usepackage[normalem]{ulem}

\newcommand{\be}{\begin{equation}}
\newcommand{\ee}{\end{equation}}
\newcommand{\ba}{\begin{eqnarray}}
\newcommand{\ea}{\end{eqnarray}}

\newcommand{\tr}{\operatorname{Tr}}

\begin{document}
	
\title{Device-Independent Quantum Key Distribution Using Random Quantum States}

\author{Subhankar Bera}
\email[Corresponding author; ]{berasanu007@gmail.com}
\affiliation{S. N. Bose National Centre for Basic Sciences, Block JD, Sector III, Salt Lake, Kolkata 700 106, India}

\author{Shashank Gupta}
\email{shashankg687@bose.res.in}
\affiliation{S. N. Bose National Centre for Basic Sciences, Block JD, Sector III, Salt Lake, Kolkata 700 106, India}
\affiliation{QuNu Labs Pvt. Ltd., M. G. Road, Bengaluru, Karnataka 560025, India}

\author{A. S. Majumdar}
\email{archan@bose.res.in}
\affiliation{S. N. Bose National Centre for Basic Sciences, Block JD, Sector III, Salt Lake, Kolkata 700 106, India}

\date{\today}
\begin{abstract}

We Haar uniformly generate random states of various ranks and study their performance in an entanglement-based quantum key distribution (QKD) task. In particular, we analyze the efficacy of random two-qubit states in realizing device-independent (DI)  QKD. We first find the normalized distribution of entanglement and Bell-nonlocality which are the key resource for DI-QKD for random states ranging from rank-1 to rank-4. The number of entangled as well as Bell-nonlocal states decreases as rank increases. We observe that decrease of the secure key rate  is more pronounced   in comparison to that of the quantum resource with increase in rank. We find that the pure state and Werner state provide the upper and lower bound, respectively, on the minimum secure key rate of all mixed two qubit states  possessing the same magnitude of entanglement  under general as well as optimal collective attack strategies. 

\end{abstract}

\maketitle
\section{Introduction}
Quantum mechanics offers safe encryption solution  based on fundamental laws (\textit{Quantum Cryptography}) rather than on computation difficulty (\textit{Rivest-Shamir-Adleman algorithm}) \cite{Gisin02}. Quantum key distribution (QKD) is the most celebrated protocol in quantum cryptography \cite{Feihu20}. There is another approach called \textit{post-quantum cryptography} which uses conventional cryptography to develop alternative public key encryption schemes that are hard even for a quantum computer to break \cite{Daniel17}. However, these are secure against the known quantum attacks, whereas security of QKD protocols is, in principle, independent of all future advances in computational power or algorithm.

There are two distinct classes of QKD protocols in literature: (a) prepare and measure schemes \cite{BB84,Bennet92}, and (b) entanglement based schemes \cite{ekert}. In prepare and measure schemes, one party prepares the quantum state and encodes the key information which is then transmitted to the other party who decodes this by performing specific measurements. The security is based on the \textit{no-cloning principle} \cite{Wootters92}. On the other hand, the entanglement based scheme uses the entanglement between the parties to share the key. The security is based on monogamy relations \cite{Pawlowski10,Pramanik}. 

Device imperfections and implementation loopholes in realistic QKD setups can compromise the security of any QKD protocol. However, device-independent QKD protocols based on entanglement remove such concern over imperfections by demonstrating QKD using uncharacterized devices \cite{Acin07, vazirani}. Security can be checked using classical constraints on correlations between the parties via Bell's inequalities, though it has been shown recently,  that  violation of Bell-CHSH inequality is not sufficient for secure QKD \cite{Jaskaran, Farkas21}. 
Device-independence allows 
QKD with uncharacterized devices \cite{Acin07,Jan20,Metger21,Nadlinger21,Xu21}. Its security has been proven
effective against collective attacks \cite{Pironio09, Ferenczi12}. On a different front, device-independent quantum secure direct communication has been recently proposed \cite{LanZhou, LanZhou1, WeiZhang}.
 Moreover, several interesting works have been proposed on QKD such as long-distance continuous-variable QKD using optical fiber\cite{YZg},  twin-field QKD\cite{PJLJ,PJLJ1}, reference-frame independent QKD using coherent states\cite{PJLJ2}, and so on.

Most of the previous works on QKD have considered specific classes of 
pure states \cite{Horodecki08}. In experiments, generation and maintenance of perfect pure states is challenging because of environmental decoherence. This results in the natural creation of  mixed states that need to be studied to get a complete picture. Two-qubit and qutrit pure states have been thoroughly studied and their performance in entanglement based QKD has been properly analyzed \cite{BB84, Bennet92,ekert,Gisin02,Kas03}. On  the other hand, limited results are known for mixed states \cite{Gisin02,Bennet92} because of multiple state parameters giving rise to multivariate optimisation problems. In this work our aim is to
investigate the performance of two-qubit mixed states of different ranks in entanglement based QKD.

Random states appear naturally in any experimental system.  They
not only arise naturally in chaotic processes, but
 can be generated also in a systematic manner based
on randomness in the outcome of quantum measurements \cite{Ratul20}. Moreover, against the intuition of observing random behavior, it has been found that random states exhibit some universal features. Examples include the performance of random states for certain communication tasks wherein it has been shown that
the dense coding capacity as well as the teleportation fidelity decrease with increase in the
rank of randomly generated states \cite{Rivu21}.

Randomly generated density matrices \cite{Vivien02,Waldemar19,Gross09,Soorya19} provide a vital tool for studying the trends of typical states in state space.
Random states were instrumental in disproving a
long-standing conjecture in quantum information theory regarding additivity of minimal output entropy
\cite{Benjamin96}. Random states have been also utilized for constructive feedback from a non-Markovian noisy environment \cite{Rivu20}. Recently,  advantage of employing two random key basis instead of one in device independent(DI)-QKD has
been demonstrated  \cite{Schwonnek21}. Some recent interesting works have been proposed on DI-QKD such as rate–distance limit of DI-QKD\cite{YMX01}, photonic demonstration of DI-QKD\cite{WZL01}, and so on. The above studies motivate us to
explore whether some universal understanding of DI-QKD tasks
could be obtained using random states.

In the present work we investigate the performance of Haar uniformly generated random
states in entanglement based QKD tasks. In particular, we estimate the average secure key rate of states having
different ranks in DI-QKD.
We first inspect the resourcefulness of the generated random state by quantifying its entanglement and Bell-nonlocality. Our results  show that the efficacy of DI-QKD in terms
of the secure key rate decreases with the
increase of rank of the random state. 
We further demonstrate that for mixed two-qubit states of any rank possessing the
same magnitude of entanglement,
the secure key rate of DI-QKD  lies  between the secure key rate of a pure state and that of a Werner state under general as well as optimal collective attack strategies.   

The paper is organised in the following way. In Sec.(\ref{Prelim}), we recapitulate the generation of random states of different ranks with the
aim of utilizing them as resource for DI-QKD. In Sec.(\ref{DI}), we present the device-independent QKD scenario under consideration and provide our analysis for the resourcefulness of the randomly generated states in terms 
of Bell-nonlocality, as well as their secure key rates.  Finally, we 
present a summary of our results in Sec.(\ref{Conclusion}).

\section{Preliminaries}
\label{Prelim}

Let us first briefly describe the procedure to generate random states. 
 We randomly simulate complex numbers from a Gaussian distribution with mean $0$ and standard deviation unity, denoted \(G(0,1)\). This ensures that the measure is Haar uniform.

\textit{Pure states:} Two-qubit pure states are then randomly generated using four such random complex numbers.
\begin{equation}
\label{R1}
    \ket{\psi_1}=\sum_{ij}c_{ij}\ket{i}\otimes\ket{j}
\end{equation}
 Here, $\ket{i}, \ket{j} \in \{\ket{0}, \ket{1}\}$
form the computational basis of the first and second qubit respectively.

\textit{Mixed states:} Random two-qubit mixed states of various ranks are generated from an appropriate pure state in a product Hilbert space by partial tracing of the suitable subsystem. 

\textit{Rank-2:} Mixed two-qubit density matrices of rank-2 are generated from random tripartite pure states  in $2\otimes 2 \otimes 2$ by tracing out any one of the three qubits \cite{Rivu20, Rivu21}.
\begin{equation}
\label{R2}
    \ket{\psi_2}=\text{Tr}_i\Big[\sum_{i,j,k =0,1}c_{ijk}\ket{i}\otimes\ket{j}\otimes\ket{k}\Big]
\end{equation}

\textit{Rank-3:} Mixed two-qubit density matrices of rank-3 are generated from random tripartite pure states  in $3\otimes 2 \otimes 2$ by tracing out the qutrit \cite{Rivu20, Rivu21}.
\begin{equation}
    \ket{\psi_3}=\text{Tr}_i\Big[\sum_{i =0,1,2}\sum_{j,k=0,1}c_{ijk}\ket{i}\otimes\ket{j}\otimes\ket{k}\Big]
\end{equation}

\textit{Rank-4:} Mixed two-qubit density matrices of rank-4 are generated from random quadripartite pure states  in $2\otimes 2 \otimes 2 \otimes 2$ by tracing out any two of the four qubits \cite{Rivu20, Rivu21}.
\begin{equation}
    \ket{\psi_4}=\text{Tr}_{ij}\Big[\sum_{i,j,k,l =0,1}c_{ijkl}\ket{i}\otimes\ket{j}\otimes\ket{k}\otimes\ket{l}\Big]
\end{equation}

Next, let us recapitulate the quantum resources that are relevant for the present study.  Quantum mechanics offers several non-classical resources that give advantage in different communication tasks. Here, we are interested in the following resources: 

\textit{Entanglement:}  Entanglement of any two-qubit state can be quantified using Negativity and Logarithmic Negativity. Using Eq.($1$) to ($4$) we generate rank-$1$ to rank-$4$ random states respectively, and we take partial transpose of those numerically generated states and determine the eigenvalues $\{\lambda_1,\lambda_2,\lambda_3,\lambda_4\}$.
Logarithmic Negativity is defined as
$LN = \log_2(2N+1)$ (
where, $N(=|\sum_j\lambda_j|)$ is Negativity and $\lambda_j$ are the negative eigenvalues of the partially transposed state).

\textit{Bell-Nonlocality:} For a given random two-qubit state $\rho_{AB}$
the maximum value of Bell-Nonlocality that can be achieved for optimal measurements is $2\sqrt{\lambda_1^2+\lambda_2^2}$.
Here, $\lambda_1$ and $\lambda_2$ are the two largest singular values of the correlation matrix T ($t_{ij} = \text{Tr}[(\sigma_i \otimes \sigma_j).\rho_{AB}]$) and $\sigma_{i(j)}$ are the Pauli matrices.

We study the performance of randomly generated states in DI-QKD tasks.  
Our entire calculations and
analysis are based on $10^6$ Haar uniformly generated states for each rank. The distribution
of states is quantified in terms of the following parameters, as defined below.

 For a given rank of random state, the normalized distribution of quantum resource is defined as the ratio between the number of states having an amount of QR, i.e., $a\leq Q_c \leq b$, with $Q_c$ being the measure of quantum correlation (entanglement or Bell nonlocality) and the total number of generated random states. Mathematically,
\begin{equation}
\mathbb{F}_D^n = \frac{\text{Number of states with } Q_c \in [a,b]}{N_0}
\label{norm}
\end{equation}
 with $N_0$ being the total number
of simulated states. Here, `n' stands for normalized and `D' stands for distribution.  $Q_c$ denotes logarithmic negativity  and violation of Bell-CHSH inequlity, in case of entanglement and Bell-nonlocality, respectively. We  divide the range of $Q_c \in (0,1]$ in 10 parts to determine the normalized distribution of quantum resource in simulated random states.  The normalized distribution of entanglement is given by
\begin{equation}
E_{nD} = \frac{\text{Number of states with} LN \in [a,b]}{N_0}
\label{Ee}
\end{equation}
Similarly, the normalized distribution of Bell-nonlocality is defined as
\begin{equation}
N_{nD} = \frac{\text{Number of states with Bell violation} \in [a,b]}{N_0}
\label{BV_normalized}
\end{equation}
The mean distribution of quantum resource is the ratio between the total number of quantum resourceful state and the total number of generated random states for a fixed rank, given by
\begin{equation}
\mathbb{F}_D^m = \frac{\sum \mathbb{F}_D^n}{N_0}
\label{meanQR}
\end{equation}
where we have summed over the entire range of $a$ and $b$. Here, `m' stands for mean distribution. This quantity represents the fraction of resourceful states. We consider Bell-nonlocal correlations as quantum resource and analyse the mean distribution of the Bell-nonlocality for the randomly simulated random states as follows:
\begin{equation}
N_{mD} = \frac{\text{Number of states violating Bell-inequality}}{N_0}
\label{mean_BV}
\end{equation}

We investigate the performance of the random states based on the above mentioned quantities. As observed from previous studies \cite{Rivu21}, the number of resourceful state decreases as the rank increases. For a particular rank, the fraction of Bell-nonlocal states is lower than that of entangled states,
exemplifying the hierarchy of these correlations for a large number of random states \cite{Wiseman07}.

\begin{figure}[ht]
	\resizebox{9cm}{7.5cm}{\includegraphics[]{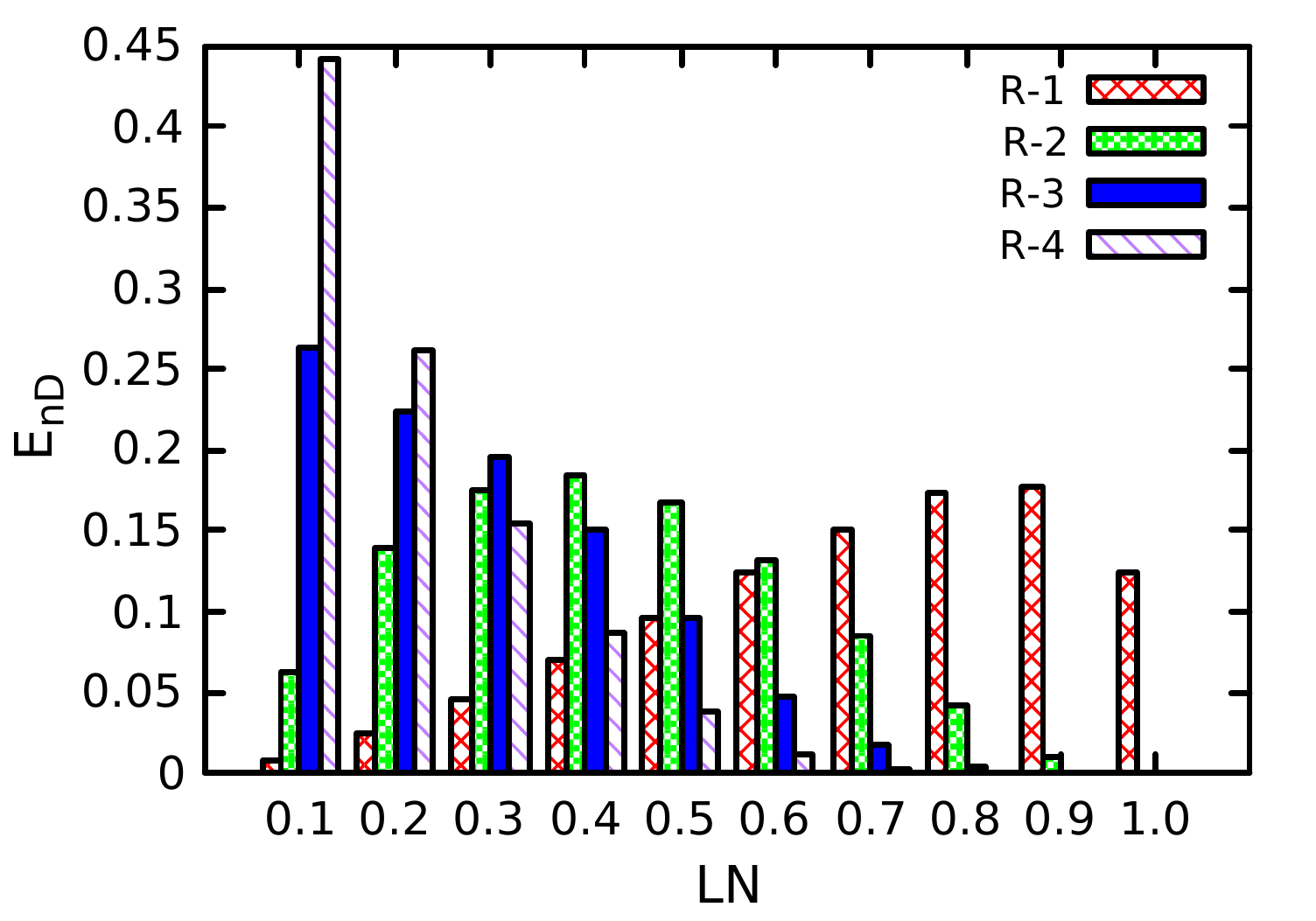}}
	\caption{(Coloronline) Normalized distribution of entangled($E_{nD}$) random two-qubit states (vertical axis) against Logarithmic negativity (LN) (horizontal axis). We mentioned only the upper value of the LN (b) in the horizontal axis for brevity of notation. Thus, 0.1 denotes the range  (0,0.1].}
\label{normalised}
\end{figure}

In Fig. (\ref{normalised}) we plot the normalized distribution of entangled random
two-qubit states against Logarithmic negativity. 
As shown in fig. (\ref{normalised}), a large fraction of simulated pure states 85\% have higher value of logarithmic negativily (0.5 and above), whereas mixed state have percentage 43.8, 16.6, 5.5 respectively for rank-2, 3, 4 states that have logarithmic negativity 0.5 and above. This implies that as the rank of the state increases, its tendency to have higher value of entanglement decreases. 
We observed that the quantum resourcefulness of the state decreases as the rank increases. The rest of the paper attempts to answer whether similar behavior is observed in the entanglement based quantum key distribution task. Specifically, we address the effect of rank and QR of the random state on its performance in DI-QKD.

\section{Bell-nonlocality and secure key rate of DI-QKD}\label{DI}

\begin{figure}[ht]
	\resizebox{8.5cm}{5cm}{\includegraphics[]{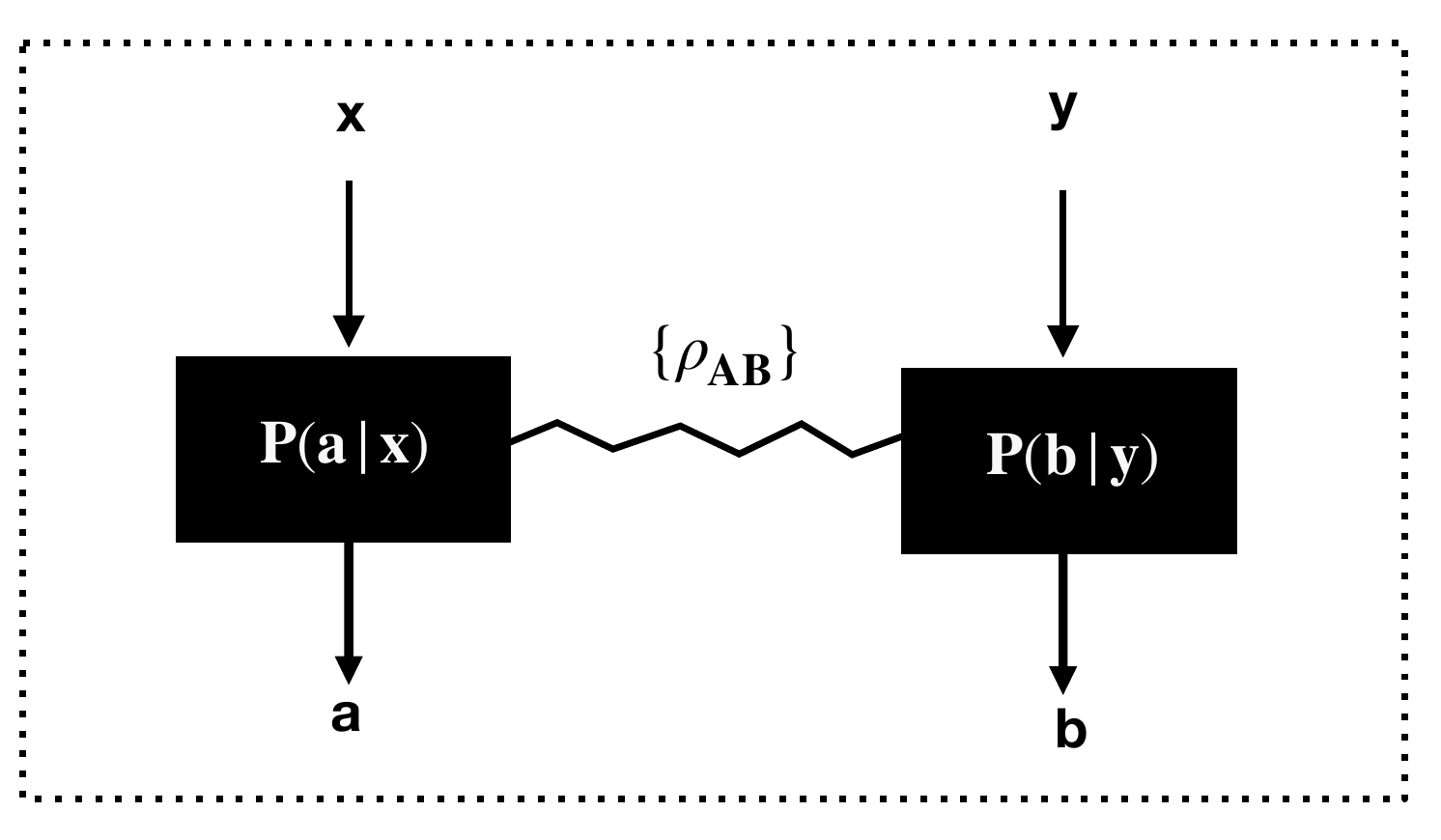}}
	\caption{(Coloronline) The device-independent quantum key distribution task.}
\label{DIfig}
\end{figure}

Let us first briefly recapitulate the protocol of DI-QKD.
Consider the two uncharacterized parties Alice and Bob sharing a bipartite entangled state $\rho_{AB}$ in $\mathbb{C}^2 \otimes \mathbb{C}^2$ as shown in fig.(\ref{DIfig}).  The two parties want to establish a secure key. For this, each of them perform dichotomic measurements in two mutually unbiased measurement bases (MUBs) and  get two outcomes. Alice performs
measurement of the observables randomly chosen from
the input $x \in \{0, 1\}$ and gets the outcome $a \in \{0, 1\}$. Similarly, Bob randomly chooses the input measurement $y \in \{0,1\}$ and gets the outcome $b \in \{0, 1\}$. In the  post-processing stage, both the parties publicly compare their input measurements
and keep only those outcomes for which their inputs are correlated.

 Our protocol is similar to E91 protocol \cite{ekert}. In a DI-QKD 
protocol, the devices are untrusted. The security is guaranteed by checking 
Bell-inequality violation from the measurement statistics. The basic steps of our DI-QKD protocol are as follows:

\textit{Quantum state preparation:} Alice generates a pair of entangled photons at her lab (random two-qubit state). She keeps one of the entangled photons and sends  the other to Bob's lab through a quantum channel.

\textit{Quantum measurement:} Alice  performs measurement of the observable randomly chosen from the input $x \in \{0, 1\}$ and gets the outcome $a \in \{0, 1\}$. Similarly, Bob randomly chooses the input measurement $y \in \{0, 1\}$ and gets the outcome $b \in \{0, 1\}$. During post-processing stage, Alice and Bob keep the cases when their inputs are correlated and discard all other cases.

\textit{Bell-inequality violation:} Alice and Bob use a fraction of inputs and outputs to ensure the Bell-inequality violation. Those cases are also discarded as the output values are disclosed.

\textit{Secure symmetric key generation:} Alice and Bob perform bi-directional error correction on their output values and perform privacy amplification on the corrected keys depending upon the information disclosed (function of the quantum bit error rate (QBER)) and Eve's attacking strategy. The final keys are the secured symmetric keys. The equality can be verified using a family of universal hash functions.


Let us determine the secret key rate under different Eve's attack strategies. In the ideal scenario with no attack, Alice and Bob are left with perfectly identical keys. However, imperfections in state preparation, transmission, measurement processes and eavesdropping can yield differences in their key strings. Alice and Bob can estimate the error rate after comparing a small portion of their secure key. Formally,  QBER for a given state $\rho_{AB}$ is defined as the average mismatch between the outcomes of Alice and Bob. Let us denote Alice's two MUBs as $\{|x^{\alpha}_a\rangle\}_{a=0}^1$ (for $\alpha$ $\in$ (0,1)) which
are correlated to Bob’s MUBs $\{|y^{\alpha}_b\rangle\}_{b=0}^1$ (for $\alpha$ $\in$ (0,1)). The perfect correlation between Alice and Bob would implies that
Alice and Bob perform measurements in the
same basis and when Alice’s outcome is $|x_a^1\rangle$,
Bob’s outcome must be $|y_b^1\rangle$. In the non-ideal scenario, there can
be non-zero probability of observing $|x_a^1\rangle$ in Alice’s subsystem
and $|y_b^1\rangle$ in Bob’s subsystem where $a \neq b$. Hence, the QBER which is an average of all these mismatch probabilities can be
expressed as
\begin{align}
    QBER &= \frac{1}{2}\sum_{\alpha=0}^1 \sum_{a\neq b=0}^1 \langle x_a^{\alpha} y_b^{\alpha}|\rho_{AB}|x_a^{\alpha} y_b^{\alpha}\rangle \nonumber \\
    &= \frac{1}{4}(2 - |\lambda_1| - |\lambda_2|)
    \label{QBER_r}
\end{align}
where $\lambda_1$ and $\lambda_2$ are the two largest singular values of the correlation matrix T ($t_{ij} = \text{Tr}[(\sigma_i \otimes \sigma_j).\rho_{AB}]$) each of which is bounded from above
by 1. 

The security of entanglement based QKD necessarily requires the demonstration of nonlocal correlations. So, for example, violation of Bell-CHSH inequality is required for the security of a DI-QKD since, none of the two parties are trusted in this scenario. Note that the violation of the Bell-CHSH inequality is the necessary criterion and not sufficient \cite{Farkas21}, and hence,  there are states that violate the Bell-CHSH inequality but still are not useful for the task of key distribution. 

Note that  for a given two-qubit state $\rho_{AB}$
the maximum value of Bell-CHSH inequality that can be achieved for optimal measurements is $2\sqrt{\lambda_1^2+\lambda_2^2}$ (say, S).
Using Eq.(\ref{QBER_r}) QBER can be written in terms of Bell-Nonlocality ($S$) as
\begin{align}
    QBER &= \frac{1}{2}\bigg(1-\sqrt{\frac{S^2}{16}+\frac{1}{2}|\lambda_1||\lambda_2|}\bigg)
    \label{Q_S}
\end{align}
From this above equation we can see that with increase of Bell-Nonlocality ($S$), the QBER may decrease.
The security proof provides a bound on the rate at which Alice and Bob can extract a secure key. The rate at which unconditionally secure key against Eve's attacks can be extracted is given by
\begin{equation}
    r(\rho_{ABE})=I(A:B)-I (A:E)
\end{equation}
where, $\rho_{ABE}$ is the joint state between Alice, Bob and Eve and I is the Holevo quantity or the quantum mutual information. Usually, the joint state $\rho_{ABE}$ is not known to Alice and Bob. So, the key rate is calculated from the QBER estimation after the error correction algorithm and the effective state after the postselection (sifting etc.) is given by
\begin{equation}
    \xi(\rho_{AB})=\sum_{u}p(u)\rho_{XYE}^u \otimes |u\rangle \langle u|
    \label{effective_state}
\end{equation}
The effective key rate is then,
\begin{equation}
    \Bar{r}(\xi(\rho_{AB}))=\mathbb{I}(\xi(\rho_{AB}))-\mathbb{I}'(\xi(\rho_{AB}))
    \label{effective_keyrate}
\end{equation}
where, $\mathbb{I}(\xi(\rho_{AB})) = \sum_u p(u)I_u(X:Y)$ and $\mathbb{I}'(\xi(\rho_{AB})) = \sum_u p(u)I_u(X:E)$. Eve has the freedom to choose any attack, if it creates a state $\rho_{AB}$ contained in the set of all bipartite states $\{\rho_{AB}\}$ that are compatible with the measurement outcomes $p(a,b|x,y)$, and  have a given reduced state $\rho_A$. The minimum secure key rate under such assumption is
\begin{equation}
    r_{\text{min}}=\underset{\{\rho_{AB}\}}{\text{inf}}\Bar{r}(\xi(\rho_{AB}))
    \label{r_mineqn}
\end{equation}
Since the global state shared between Alice, Bob and Eve are not known,  the secure key rate can be determined as a function of the QBER  using Eq. (\ref{effective_state}), (\ref{effective_keyrate}) and (\ref{r_mineqn}).

{\it Secret key rate under collective attacks(CA):}
In the case of collective attacks, the eavesdropper applies the same attack on each system of Alice and Bob. Here the minimum secure key rate is a function of QBER(Q) and S. The minimum secure key rate is given by \cite{Pironio09}
\begin{equation}
    r_{\text{Cmin}} \geq 1 - h(Q) -h\bigg(\frac{1 + \sqrt{(S/2)^2-1}}{2}\bigg)
    \label{DIrateC}
\end{equation}
Where h is binary entropy and S (= $2\sqrt{\lambda_1^2+\lambda_2^2}$) is the Bell-CHSH violation.

 {\it Secret key rate under optimal symmetric collective attacks(OSCA):}
For the case of optimal symmetric collective attacks (attack optimised over the symmetries of the protocol, state and measurements of the communicating parties) by the eavesdropper in entanglement assisted protocols for two-qubit states with two measurement settings per qubit, the minimum secure key rate is given by \cite{Ferenczi12}
\begin{equation}
    r_{\text{Smin}} = 1 + 2(1 - Q) \text{log}_2(1 - Q) + 2Q \text{log}_2 Q 
    \label{DIrate}
\end{equation}

We have two separate conditions for the security of a DI-QKD protocol.
One being $r_{\text{C(S)min}} > 0$ for a secure key to be distilled while
the second is the requirement that the underlying entangled state violates the Bell-CHSH inequality. While it can be seen that there exist no states with non-vanishing secure
key and no Bell-CHSH violation, there do exist states
which show Bell-CHSH violation but have vanishing secure key.

\begin{figure}[ht]
	\resizebox{9cm}{8cm}{\includegraphics[]{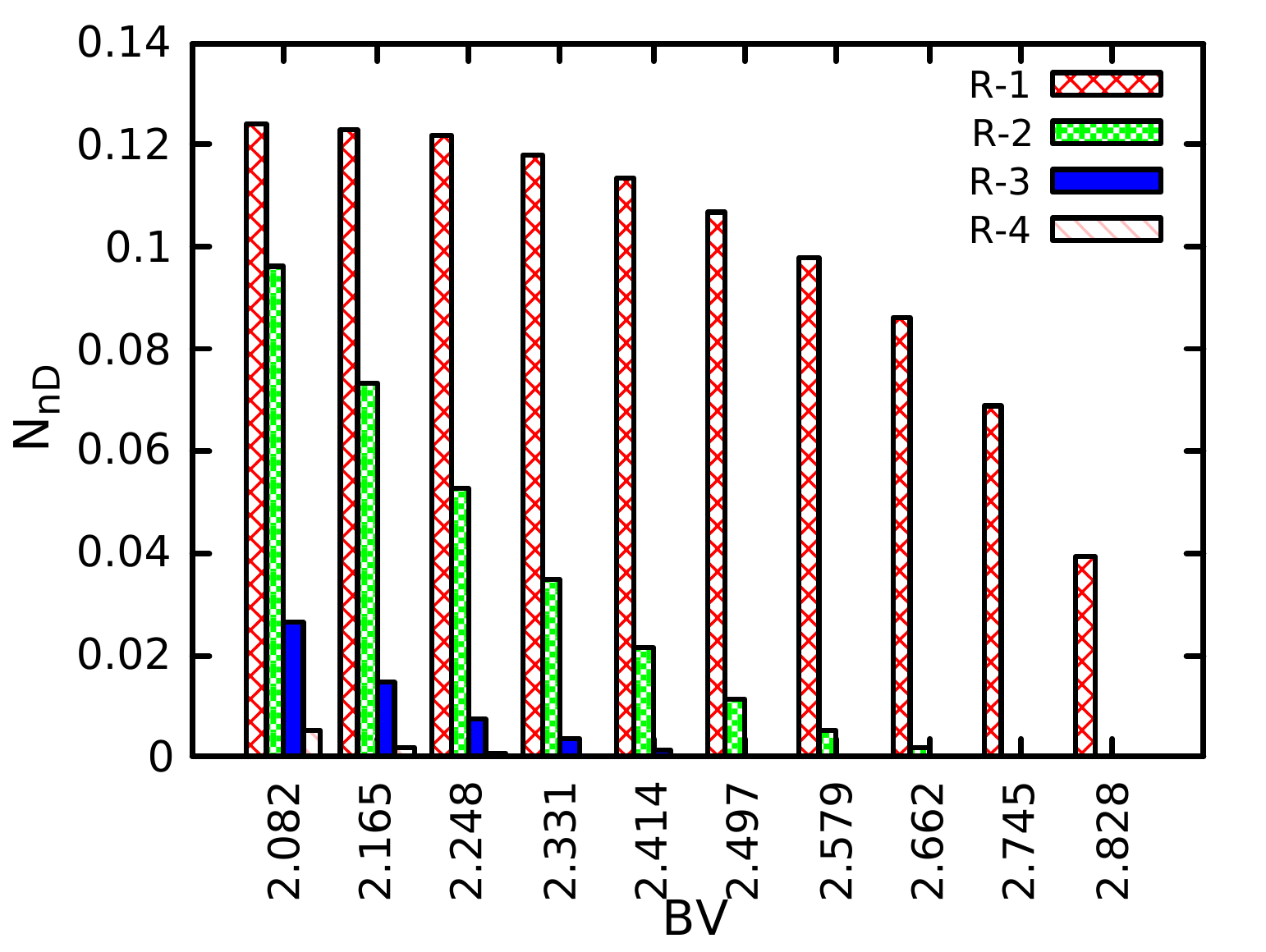}}
	\caption{(Coloronline) Normalized distribution of Bell nonlocal($N_{nD}$) random two-qubit states (vertical axis) against the violation of the Bell-CHSH inequality (BV) (horizontal axis). We mention only the upper value of the Bell's inequality violation in the horizontal axis for brevity of notation. Thus, 2.082 denotes the range (2,2.082].}
\label{normalised_bell}
\end{figure}

We now study the behaviour of Bell-nonlocal correlations and minimum secure key rate of random states in DI-QKD. In particular, we first analyse the normalised distribution of Bell-nonlocal correlations (Eq.(\ref{norm})) as shown in fig. (\ref{normalised_bell}). It is seen that
the tendency of a random state to achieve large value of Bell-CHSH inequality (2.5 and above) decreases with increasing rank. We find 39.9, 1.9, 0.05, and 0.001 
to be the respective percentage of the simulated rank-1, 2, 3 and 4 states that achieve Bell-inequality value of 2.5.

\begin{figure}[ht]
	\resizebox{8.5cm}{7.5cm}{\includegraphics[]{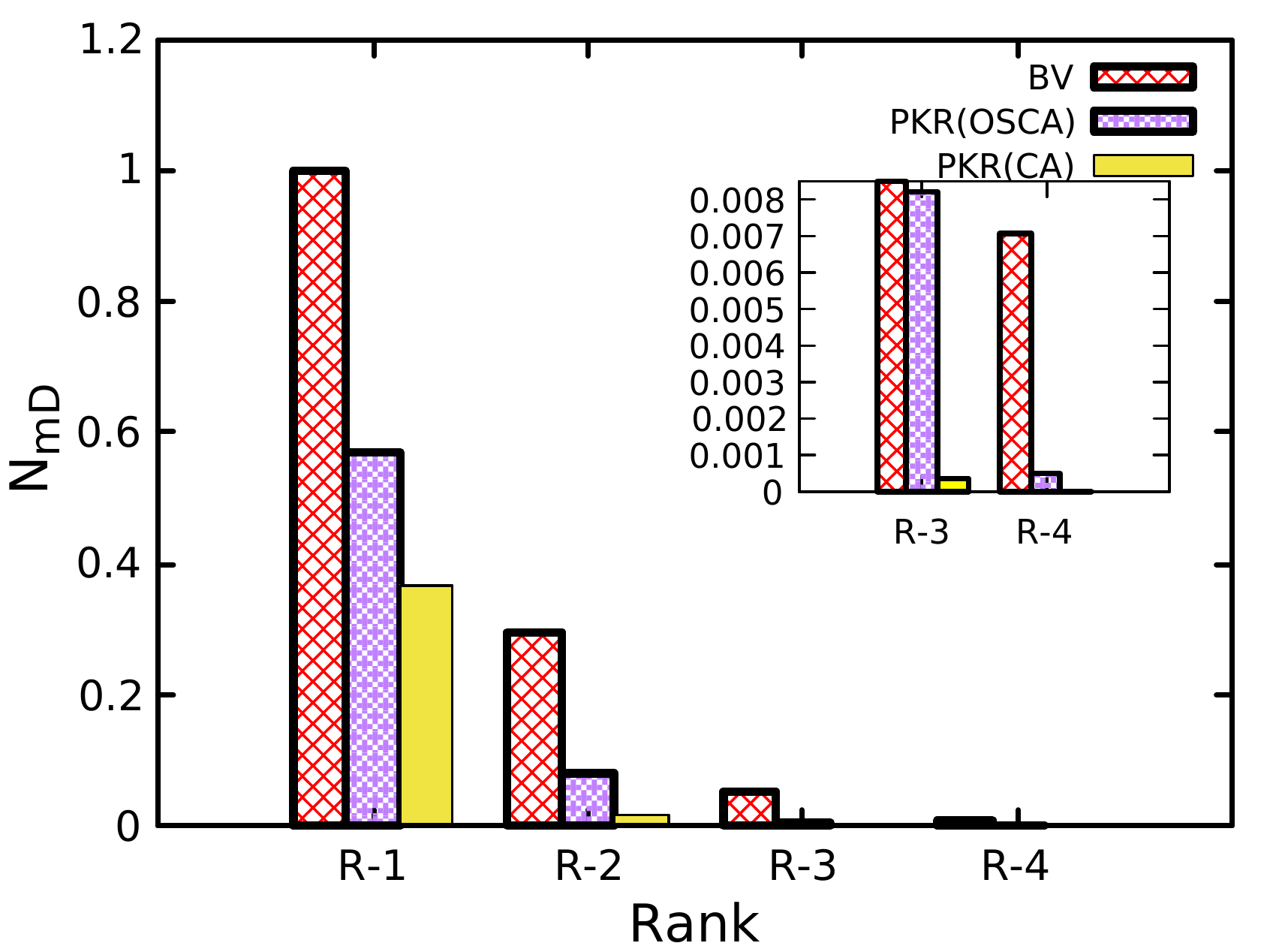}}
	\caption{(Coloronline)  The mean distribution of Bell-nonlocal ($N_{mD}$) random two-qubit states as well as the fraction of random two-qubit states that have  minimum secure positive key rate (PKR), $r_{\text{C(S)min}}$ for the given rank of the states under optimal symmetric collective attacks(OSCA) and collective attacks(CA) for different rank of the random two-qubit state. }
\label{before_hist_TF}
\end{figure}

We next perform a comparative study of  the mean distribution of $r_{\text{C(S) min}}$ and Bell-nonlocality  of all four ranks. The fraction of random states that have non-zero value of the secure key rate is given by
\begin{equation}
\mathbb{F}_D^r = \frac{N^r}{N_0}
\label{mean_r}
\end{equation}
where $N^r$ is the number of states that have non-zero value of secure key rate in DI-QKD. 
Using Eqs. (\ref{meanQR}) and (\ref{mean_r}), we plot the respective
distributions for all four ranks in Fig.(\ref{before_hist_TF}).

It is seen that the  number of randomly simulated states that are Bell-nonlocal as well as the states which provide positive minimum secure key rate decreases with the increase of the rank of the states. The percentages of states that are Bell-nonlocal and give positive secure key rate are 56.8, 8.2, 0.70 and 0.05 for rank-1, 2, 3 and 4, respectively under optimal symmetric collective attacks. Similarly, under collective attacks, the  percentages are 36.8, 1.6, 0.04 and 0.001 for rank-1, 2, 3 and 4, respectively, under collective attacks. Hence, the number of states giving positive secret key rate under general collective attack are less than that under optimal symmetric collective attacks. 

Note that the respective percentage of Bell-nonlocal states are higher in
both cases. This again implies that all Bell-nonlocal states are not suitable for  DI-QKD, reinforcing a similar claim in 
a recent work  \cite{Farkas21}. Moreover, the rate of decrease in $r_{\text{C(S)min}}$ is more prominant than Bell-nonlocality implying that higher rank Bell-nonlocal states are less useful for DI-QKD. The number of states that are Bell non-local and have positive secure key rate under general collective attacks is less in comparison to that in the optimal symmetric collective attack for every rank. This behaviour is expected because in 
the general collective attack strategy, Eve has the freedom to devise a strategy to maximise  mutual information whereas in the optimal symmetric attack, the quantum protocol, state and measurement symmetries put constraints over the strategy of Eve. This constrains  Eve's mutual information and relaxes the secure key rate requirements.

 For a given rank of the random state, 
the average secure key rate is given by the ratio of the sum of the secure key rate of the simulated states to the number of states that have non-zero value of the secure key rate, as
\begin{equation}
    \Bar{r}=\frac{\sum_i r_i}{N'}
    \label{aver_key}
\end{equation}
where, $r_i$ is the secure key rate of the $i^{\text{th}}$ state and $N'$ is the total number of states that have non-zero value of the secure key rate.
The average key rate computed using Eq. (\ref{aver_key}) in DI-QKD under optimal symmetric collective attacks is 0.36, 0.15, 0.09 and 0.07 for rank-1, 2, 3 and 4 states,  whereas, under collective attacks, the average key rate is 0.34, 0.14, 0.09 and 0.06 for rank-1, 2, 3 and 4 states, respectively, as shown in Table-(\ref{DI_table}). The average key rate in both situations where Eve does a general or optimal collective attack decreases with increasing rank implying that the tendency to generate positive secure key rate decreases with increasing rank. The average key rates in both the attack strategies are nearly the same for a given rank. Our entire calculations are based on $10^6$ Haar uniformly generated states for each case. We find  that a large fraction of pure states have positive value of minimum secure key rate and are Bell-nonlocal in comparison with the mixed two-qubit states. This is in accordance with a previous study \cite{Rivu20} where it was observed that large fraction of randomly generated mixed states are Bell local states. However, this is in contrast with the observation for teleportation fidelity where it was found that with increasing rank, relative number of states that are local but gives non-classical fidelity increases \cite{Rivu21}.

\begin{widetext}
\begin{center}
 \begin{table}[h!]
			\caption{Average secure key rate in DI scenario} 
			\label{DI_table}
			\centering
		\begin{tabular}{|r|r|r|r|r|r|r}
			\hline
			& \begin{tabular}[c]{@{}l@{}}No. of random states\\ that  violate the Bell-CHSH \\ inequality (among $10^6$\\ random states)\end{tabular} & \begin{tabular}[c]{@{}l@{}}No. of random states\\ that have positive\\ secure key rate under\\ OSCA\end{tabular} & \begin{tabular}[c]{@{}l@{}}No. of random states\\ that have positive\\ secure key rate under\\ CA\end{tabular} &\begin{tabular}[c]{@{}l@{}}Average secure\\ key rate(OSCA)\end{tabular} &\begin{tabular}[c]{@{}l@{}}Average secure\\ key rate\\(CA)\end{tabular}\\ \hline
			R-1 & 1000000  & 568522  &  368453 & 0.36    &  0.34 \\ \hline
			R-2 & 297642 & 82314 &  16662 & 0.15 & 0.14 \\ \hline
			R-3 & 54464 & 7060 & 423 & 0.09 &   0.09 \\ \hline
			R-4 & 8258 & 498 & 11 & 0.07 & 0.06 \\ \hline
		\end{tabular}
\end{table}
\end{center}
\end{widetext}
 
 We observe that on average, the quantum resourcefulness of the randomly generated states decreases with an increase in rank and this could be the reason that the performance of the state also decreases in the DI-QKD task as its rank increases. In particular, pure states perform better than rank-2 states, and
 in turn, rank-2 states perform better than rank-3 and rank-4 states. Interestingly, there are states of different rank which have the same value of the entanglement, but have different value of the minimum secure key rate. To
 illustrate this feature, we next perform a comparative study of pure states, general rank-2 states and Werner states.  Werner states are the simplest and most studied two-qubit mixed states that help in understanding the effect of noise on maximally entangled Bell states. 
 We determine the minimum secure key rate of these three states in terms of the negativity to show the distinction in performance  for the same value of the entanglement.

An arbitrary two-qubit
pure state in a Schmidt decomposition has the form
\begin{equation}
  \ket{\psi_p} = \cos \frac{\theta}{2}\ket{00}+\sin \frac{\theta}{2}\ket{11}
  \label{ps}
  \end{equation}
where, $\ket{0}$ and $\ket{1}$ are the eigenstates of the reduced density matrices, and eigenvalues of the local density matrices are $\cos^2 \frac{\theta}{2}$ and $\sin^2 \frac{\theta}{2}$. The negativity of the pure state is given by the
the square root of the determinant of its reduced density matrix, i.e., $\frac{\sin \theta}{2}$.

Any two qubit mixed state of rank-2 can be expressed as,
\begin{equation}
    \rho_2^2 = p_1 |\psi_1\rangle \langle \psi_1| + (1 - p_1)|\psi_2\rangle \langle \psi_2|
\label{GeneralR2}
\end{equation}

where, $|\psi_1\rangle = \alpha|0\eta_1\rangle + \beta|1\eta_2\rangle \:$, $\: |\psi_2\rangle = \alpha|0\eta_1^\perp\rangle + \beta|1\eta_2^\perp\rangle$, $|\eta_1\rangle = a|0\rangle + b|1\rangle \:$ and $\: |\eta_2\rangle = a'|0\rangle + b'|1\rangle$ with $|\eta_1^\perp\rangle$ and $|\eta_2^\perp\rangle$ being orthogonal states to $|\eta_1\rangle$ and $|\eta_2\rangle$ respectively.
The coefficients are taken to be real for simplicity and each of the states are normalised, i.e.,
$a^2 + b^2 = a'^2 + b'^2 = \alpha^2 + \beta^2 = 1 \:$ and  $\: 0\leq p_1 \leq 1$
The entanglement of state $\rho_2^2$ in Eq. (\ref{GeneralR2}) is given by
  \begin{eqnarray}
  	N_2 &=& \frac{1}{2} \left[ \sqrt{p_1^2 - x} - p_1\right], \quad \text{if } p_1 < 0.5 \label{e2}\\
 	N_2 &=& \frac{1}{2} \left[ \sqrt{(1 - p_1)^2 + x} - (1-p_1)\right],  \text{  if } p_1 > 0.5 \label{e3}
 \end{eqnarray}
where, $x = 4\alpha^2 \beta^2 (a'b - ab')^2 (2p_1 -1)$. The state parameter $p_1$ of rank-2 state (\ref{GeneralR2}) can be expressed in terms of the negativity, as
\begin{eqnarray}
p_1 &=& \frac{N^2 -  \alpha^2 \beta^2 (a'b - ab')^2}{N - 2 \alpha^2 \beta^2 (a'b - ab')^2}, \text{  if } p_1 < 0.5 \label{neg_relation}\\
p_1 &=& \frac{N(N + 1) +  \alpha^2 \beta^2 (a'b - ab')^2}{2 \alpha^2 \beta^2 (a'b - ab')^2 + N}, \text{if } p_1 > 0.5 
\label{neg_relation1}
\end{eqnarray}

Next, the two qubit Werner state is given by 
\begin{equation}
\rho_W = p|\phi^+\rangle \langle \phi^+ | + \frac{(1-p)}{4} I_4
\label{ws}
\end{equation}
where 
$|\phi^+\rangle = \frac{1}{\sqrt{2}} (|00\rangle + |11\rangle)$ with $0\leq p \leq 1$ and $I_4$ being the identity matrix in Hilbert space $\mathbb{C}^2 \otimes \mathbb{C}^2$. One can take any other maximally entangled Bell state instead of $|\phi^+\rangle$ in the expression of the Werner state but the final expression of the minimum secure key rate is same.  The negativity of the Werner state is $\frac{3p-1}{4}$.

\begin{figure}[ht]
	\resizebox{8.5cm}{7.5cm}{\includegraphics{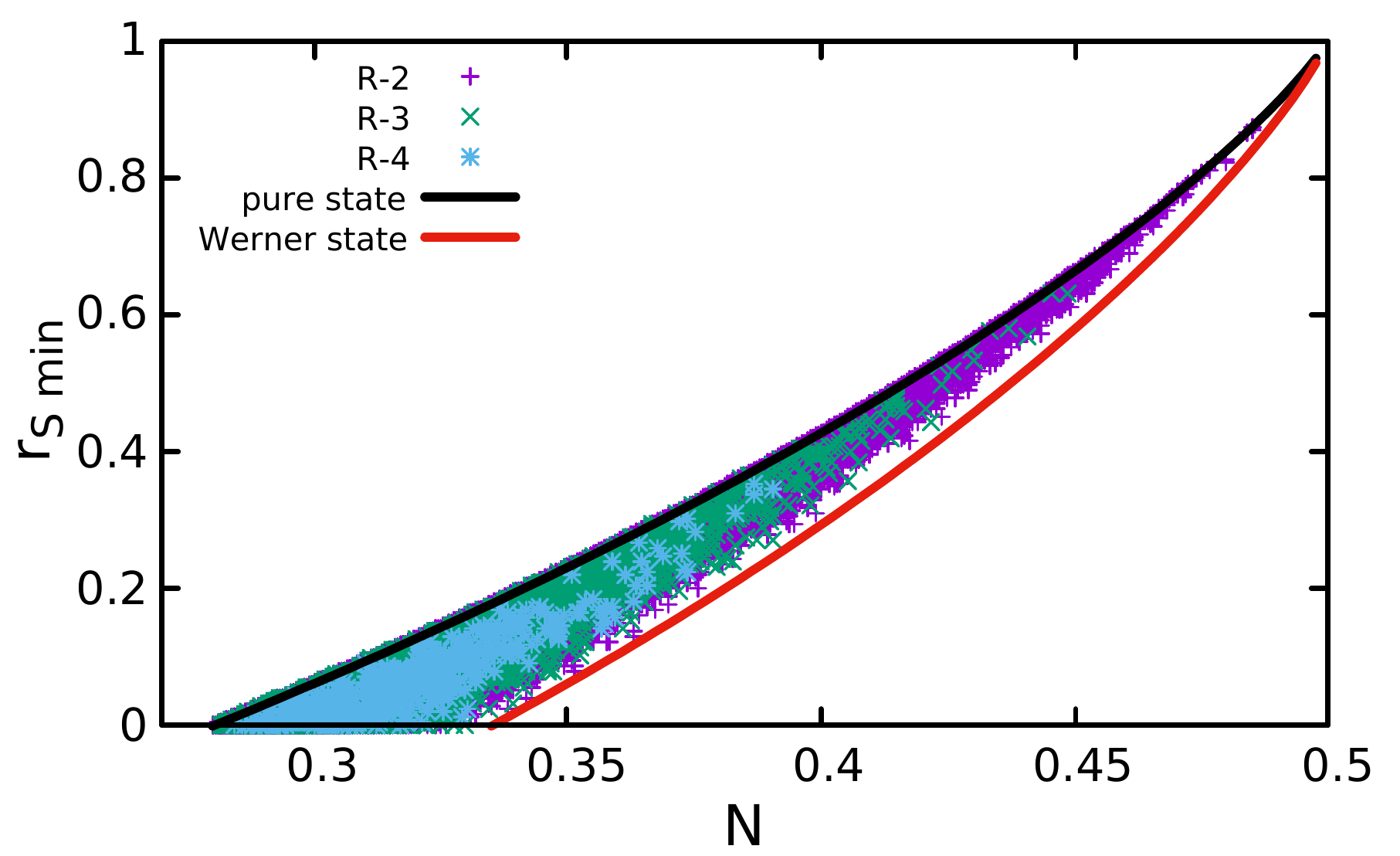}}
	\caption{(Coloronline) Minimum secure key rate of randomly generated rank-2,  rank-3, rank-4 states, pure state and the Werner state in DI-QKD are plotted 
	versus the negativity for the case of optimal symmetric collective attacks. It is clear that the pure state and the Werner state provides the upper and lower bound respectively, on the minimum secure key rate of mixed two-qubit states in DI-QKD.}
\label{theorem2DI}
\end{figure}

We now calculate the secure key rate of the rank-2  state  (\ref{GeneralR2})   in terms of negativity ($N$). Similarly, we calculate the secure key rate of the pure state and Werner state in terms of the negativity   (see Appendix (\ref{R2_DI}) for the respective expressions).
In Fig. (\ref{theorem2DI}) we plot the minimum secure key rate of the pure state, the general rank-2 state and the Werner state  in terms of negativity. From the figure it is clear that states with the same value of the negativity can have different performance ($r_{\text{C(S)min}}$) in the DI-QKD task. It can 
also be seen that the secure key rate of the rank-2 two-qubit state lies in between the secure key rate of pure state and the Werner state at same value of negativity for both categories of collective attack, i.e.,
\begin{equation}
 r_{C(S)\text{min}}(\rho_p) \geq r_{C(S)\text{min}}(\rho_2^2) \geq r_{C(S)\text{min}}(\rho_W) 
\end{equation}
where $r_{C(S)\text{min}}$ is the minimum secure key rate in our DI-QKD scenario (\ref{DIrate}).

\begin{figure}[ht]
	\resizebox{8.5cm}{7.5cm}{\includegraphics{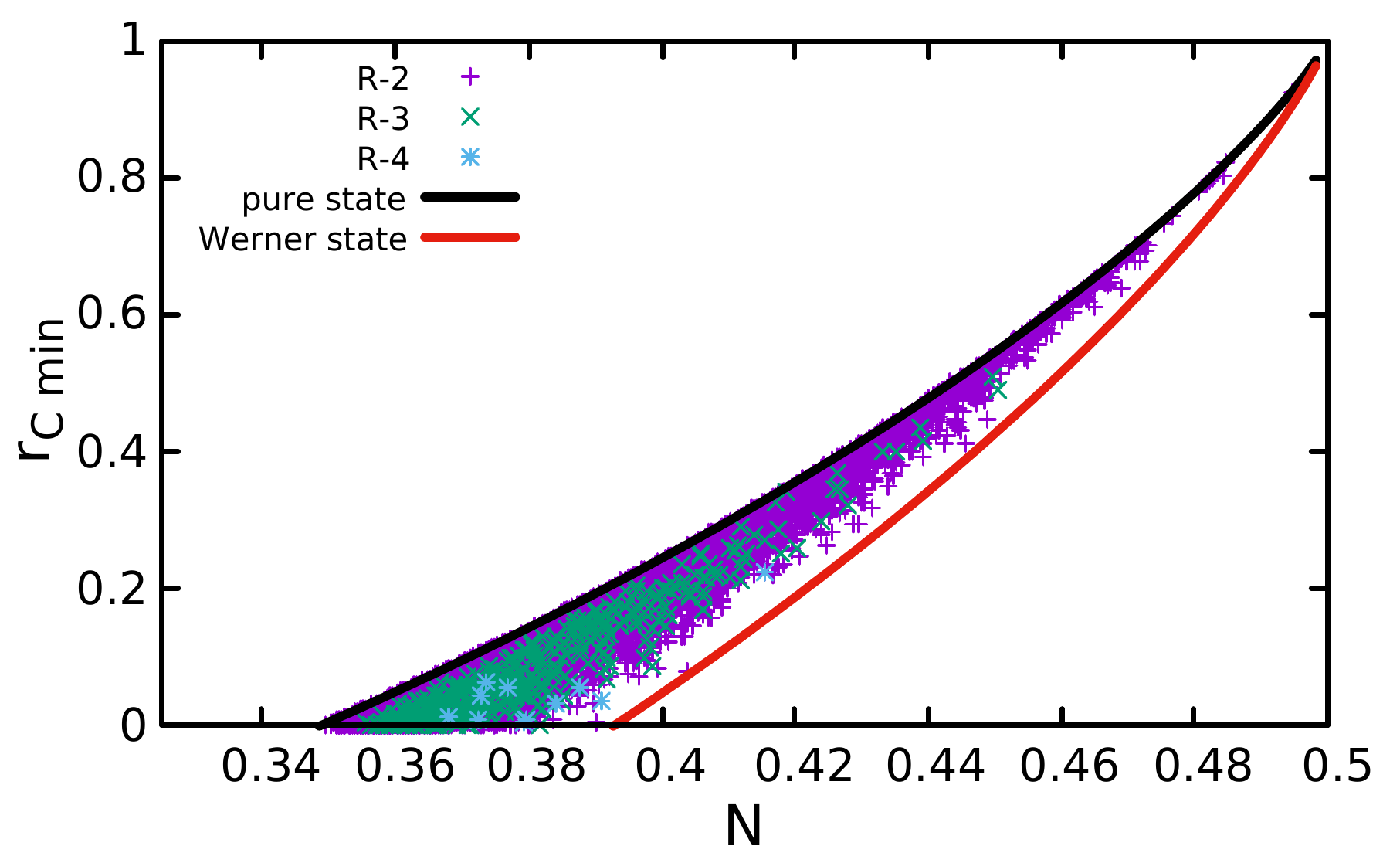}}
	\caption{(Coloronline) Minimum secure key rate of randomly generated rank-2,  rank-3, rank-4 states, pure state and the Werner state in DI-QKD are plotted 
	versus the negativity for the case of collective attacks. It is clear that the pure state and the Werner state provides the upper and lower bound respectively, on the minimum secure key rate of mixed two-qubit states in DI-QKD.}
\label{CA}
\end{figure}

 We further find numerically,  that rank-3 and rank-4 states also have the minimum secure key rate within the envelope formed by the pure state and the Werner state for the same value of negativity. From Fig.(\ref{theorem2DI}) 
 it can be observed that
 78.6\% of rank-2 states,  39.8\% of rank-3 states, and 22.7\% of rank-4 states have $r_{\text{Smin}}$ 0.1 and above under OSCA. Further, from Fig.(\ref{CA}),
 it follows that  74.6\% of rank-2 states,  28.2\% of rank-3 states, and 16.4\% of rank-4 states have $r_{\text{Cmin}}$ equals 0.1 or above under CA. All of them are inside the envelope formed by the pure state and the Werner state. Our above analysis can be summarized as following result:
 
\textbf{Result:} \textit{The secure key rate of any mixed two qubit state  in DI-QKD is lower bounded by the secure key rate of the two qubit Werner state and upper bounded by the secure key rate of the pure state possessing the same value of the negativity under general as well as optimal collective attacks by Eve.}

\section{Conclusions}\label{Conclusion}

Quantum key distribution is set to become an integral part of  modern cryptographic applications. In theory, unconditional security has been shown for the prepare and measure as well as the entanglement based schemes. However,
in practice,
perfect quantum key distribution cannot be achieved due to the presence of different decohering factors, device imperfections and implementation loopholes. Therefore, it is of prime importance to study quantum key distribution protocols using randomly generated states rather than confining to specific set of states, 
with the aim of obtining  a universal perspective. 

In this work, we have studied the secure key rate of randomly generated two-qubit states of all four ranks in entanglement based QKD. Our analysis
is based on numerical results obtained by considering $10^{6}$ states
corresponding to each rank. We first estimate the fraction of states in
each rank which are Bell-nonlocal, and the fraction of states which yield
positive secure key rate in DI-QKD under general as well as optimal collective attacks by Eavesdropper. We show that both Bell-nonlocality
and the minimum secure key rate decrease with the increase of rank in
general as well as optimal attack strategy, which is a fundamental feature of such randomly generated states. 

 From our analysis we have observed that with increasing rank the decrease in secure key rate is more pronounced compared to
Bell-CHSH violation. The ratio of the number of states that have quantum resource (entangled as well as Bell-nonlocal) as a function of rank decreases slowly in comparison to the ratio of the number of states that give positive secure key rate as a function of rank. For example, the ratio of the number of rank-3 states that are Bell-nonlocal to the number of rank-2 states that are Bell-nonlocal is 0.183, whereas the respective ratio for the number of states that give positive key rate is only 0.085 under optimal symmetric collective attacks and 0.025 under collective attack respectively. It may be noted that quantum resourcefulness is a necessary condition to obtain secure key rate. 
However, the secure key rate generation is more demanding, and hence, the number of states that give secure key rate is lesser compared to the number of resourceful states.

 Our results further show that states with the same magnitude of entanglement
can lead to different values of the secure key rate. We demonstrate
that the minimum secure key rate of all two-qubit mixed states is upper
bounded by the key rate of the pure state, and lower bounded by the key
rate of the Werner state possessing the same value of entanglement quantified
by their negativity in both optimal as well as general collective attack strategy. It would be worth studying if the above bounds can be obtained using
analytical methods. It might also be interesting to  study in future the effect
of statistical fluctuations in the number of randomly generated states on the
above bounds. Moreover, our present analysis should motivate further studies
on the resilience of random states against particular quantum attacks in
QKD protocols, as well as under other sources of error such as channel
loss and misalignment rate \cite{binghongli}.

A detailed study using random state provides a source-independent analysis and establishes an efficiency and performance profile of the quantum task under consideration. For example, the random states can give a precise idea about the performance of higher rank mixed states in tasks like quantum 
multiparty cryptography \cite{yaofu}, secure quantum secret sharing \cite{jiegu}, quantum conference key agreement \cite{zhaoli}, quantum private query \cite{AM1} and quantum secure direct communication \cite{LanZhou,LanZhou1}. This in turn should further be useful in understanding the efficiency of such tasks under decoherence. This is so because  decoherence can be modelled as a black box whose input may be a random state and the output is some different random state, in order to analyse the efficacy of employing random states in 
various quantum information protocols. 

\section{Acknowledgements} 
SB and ASM acknowledge support from the Project No. DST/ICPS/QuEST/2018/98
from the Department of Science and Technology, Government of India. SG acknowledges the support from QuNu Labs Pvt Ltd.\\
\\
{\bf Data availability statement:} The datasets generated during and/or analysed during the current study are available from the corresponding author on reasonable request.\\\\
{\bf Conflict of interest statement:} All authors declare that they have no conflict of interest with any organization or entity in the subject matter or materials discussed in this manuscript.

\appendix
\onecolumngrid

\section{Minimum secure key rate in DI-QKD}\label{R2_DI}
$\bullet$ {\bf General rank-2 state:}

The matrix form of general rank-2 state (\ref{GeneralR2}) is as follows:

\begin{equation}
\rho_2^2 = \left(
\begin{array}{cccc}
 \alpha ^2 \left(p_1 \left(a^2-b^2\right)+b^2\right) & a \alpha ^2 b \left(2 p_1-1\right) & \alpha  \beta  \left(a p_1 a'-b \left(p_1-1\right) b'\right) & \alpha  \beta  \left(b \left(p_1-1\right) a'+a p_1 b'\right) \\
 a \alpha ^2 b \left(2 p_1-1\right) & \alpha ^2 \left(p_1 \left(b^2-a^2\right)+a^2\right) & \alpha  \beta  \left(p_1 \left(b a'+a b'\right)-a b'\right) & \alpha  \beta  \left(\left(a-a p_1\right) a'+b p_1 b'\right) \\
 \alpha  \beta  \left(a p_1 a'-b \left(p_1-1\right) b'\right) & \alpha  \beta  \left(p_1 \left(b a'+a b'\right)-a b'\right) & \beta ^2 \left(p_1 \left(\left(a'\right)^2-\left(b'\right)^2\right)+\left(b'\right)^2\right) & \beta ^2 \left(2 p_1-1\right) a' b' \\
 \alpha  \beta  \left(b \left(p_1-1\right) a'+a p_1 b'\right) & \alpha  \beta  \left(\left(a-a p_1\right) a'+b p_1 b'\right) & \beta ^2 \left(2 p_1-1\right) a' b' & \beta ^2 \left(p_1 \left(b'\right)^2-\left(p_1-1\right) \left(a'\right)^2\right) \\
\end{array}
\right)
\end{equation}
Next, we compute the eigenvalues of the correlation matrix (T) of the general rank-2 state. The matrix elements of the correlation matrix are $t_{ij} = \text{Tr} [(\sigma_i \otimes \sigma_j).\rho_2^2]$. The  correlation matrix is:
\begin{equation}
    \left(
\begin{array}{ccc}
 2 \alpha  \beta  \left(2 p_1-1\right) \left(b a'+a b'\right) & 0 & 2 \alpha  \beta  \left(2 p_1-1\right) \left(a a'-b b'\right) \\
 0 & 2 \alpha  \beta  \left(b a'-a b'\right) & 0 \\
 2 \left(2 p_1-1\right) \left(a \alpha ^2 b-\beta ^2 a' b'\right) & 0 & \left(2 p_1-1\right) \left(\alpha ^2 \left(a^2-b^2\right)-\beta ^2 \left(a'\right)^2+\beta ^2 \left(b'\right)^2\right) \\
\end{array}
\right)
\label{CM_R2}
\end{equation}

The eigenvalues of the correlation matrix (\ref{CM_R2}) are:
\begin{align}
    \lambda_1 &= y \nonumber \\
    \lambda_2 &= \frac{1}{2}(1-2 p_1)\Big[\alpha^2(b^2-a^2)+\beta^2(a'^2-b'^2) - y' + \sqrt{(2ab\alpha^2 - y + \beta^2(1-2a'b') - z)(\beta^2(1+2a'b') - 2ab\alpha^2 - y + z)}\Big] \nonumber \\
    \lambda_3 &= \frac{1}{2}(1 - 2p_1)\Big[\alpha^2(b^2-a^2)+\beta^2(a'^2-b'^2)-y'- \sqrt{(2ab\alpha^2 - y + \beta^2(1-2a'b') - z)(\beta^2(1+2a'b') - 2ab\alpha^2 - y + z)}\Big]
\end{align}
where, $y=2\alpha\beta(ab'-a'b)$, $y' = 2\alpha\beta(a'b+ab')$ and $z=2\alpha\beta(a'a-bb')$.
We determine the quantum bit error rate (QBER) in DI-QKD using Eq.(\ref{QBER_r}) for the case $(ab' = a'b)$.
\begin{align}
    QBER &=\frac{1}{4}(2 - |\lambda_2| - |\lambda_3|) \nonumber\\
    &= \frac{1}{4}\Big[2-|(1-2p_1)(\alpha^2(b^2-a^2)+\beta^2(a'^2-b'^2) - y')|\Big]
\end{align}
The $r_{\text{Smin}}(\rho_2^2)$ under optimal symmetric collective attacks(OSCA), is calculated using Eq.(\ref{DIrate})
\begin{align}
    r_{\text{Smin}}(\rho_2^2(p_1,a,a',\alpha)) &= \frac{1}{2 \log (2)}\Big[\left(\left(1-2 p_1\right) \left(\alpha ^2 \left(b^2-a^2\right)+\beta ^2 \left(\left(a'\right)^2-\left(b'\right)^2\right)-y'\right)+2\right)\nonumber\\
    &\log \left(\frac{1}{4} \left(\left(1-2 p_1\right) \left(\alpha ^2 \left(b^2-a^2\right)+\beta ^2 \left(\left(a'\right)^2-\left(b'\right)^2\right)-y'\right)+2\right)\right)\nonumber \\
    &+\left(2-\left(1-2 p_1\right) \left(\alpha ^2 \left(b^2-a^2\right)+\beta ^2 \left(\left(a'\right)^2-\left(b'\right)^2\right)-y'\right)\right) \nonumber \\
    &\log \left(\frac{1}{4} \left(2-\left(1-2 p_1\right) \left(\alpha ^2 \left(b^2-a^2\right)+\beta ^2 \left(\left(a'\right)^2-\left(b'\right)^2\right)-y'\right)\right)\right)+\log (4)\Big]
\end{align}
Substituting $p_1$ in terms of $N$ using Eq.(\ref{neg_relation}) for the case $(ab' = a'b)$, we get 

\begin{align}
    r_{Smin}(\rho_2^2)=&\frac{1}{2 \log 2}\Bigg[\log 4+\log\Big(\frac{1}{4}(2-(1-2N)((b^2-a^2)\alpha^2-4 b \alpha\beta a' + \beta^2(a'^2-b'^2)))\Big)\nonumber\\
   &(2-(1-2N)((b^2-a^2)\alpha^2-4 b \alpha\beta a' + \beta^2(a'^2-b'^2)))\nonumber\\
   &+\log\Big(\frac{1}{4}(2+(1-2N)((b^2-a^2)\alpha^2-4 b \alpha\beta a' + \beta^2(a'^2-b'^2)))\Big)\nonumber\\
   &(2+(1-2N)((b^2-a^2)\alpha^2-4 b \alpha\beta a' + \beta^2(a'^2-b'^2)))\Bigg]
\end{align}

Similarly for the case of collective attacks(CA), using Eq.(\ref{QBER_r}) and Eq.(\ref{DIrateC}) we obtain the  $r_{\text{Cmin}}(\rho_p)$ in terms of negativity $N$,
 \begin{align}
    r_{\text{Cmin}(\rho_2^2)} &=  \frac{1}{\log16}\bigg[-2\log16+(2-\Omega)\log(2-\Omega)+(2+\Omega)\log(2+\Omega)\nonumber\\
    &+2(1+\sqrt{\Delta-1})\log(1+\sqrt{\Delta-1})+2(1-\sqrt{\Delta-1})\log(1-\sqrt{\Delta-1})\bigg]
 \end{align}
 
Where, $\Omega=(1-2N)((b^2-a^2)\alpha^2-4 b \alpha\beta a' + \beta^2(a'^2-b'^2)),$\\
$\Delta=2(1-2N)^2\bigg((2ab\alpha^2+\beta^2+2\alpha\beta bb'-2a'(a\alpha\beta+\beta^2 b'))
(-2ab\alpha^2+\beta^2-2bb'\alpha\beta+2a'(a\alpha\beta+b'\beta^2))\\
+((b^2-a^2)\alpha^2+((a')^2-(b')^2)^2\beta^2))\bigg).$

We vary the state parameters in the step size of 0.01 to numerically determine the  minimum secure key rate as function of the negativity ($N$).

$\bullet$ {\bf General pure state:}

The matrix form of a general pure state (\ref{ps}) is as follows:
\begin{equation}
\rho_p = \left(
\begin{array}{cccc}
 \cos^2 \frac{\theta}{2} & 0 & 0 & \frac{\sin{\theta}}{2} \\
 0 & 0 & 0 & 0 \\
 0 & 0 & 0 & 0 \\
\frac{\sin{\theta}}{2} & 0 & 0 & \sin^2 \frac{\theta}{2} \\
\end{array}
\right)
\end{equation}
The correlation matrix  $t_{ij} = \text{Tr} [(\sigma_i \otimes \sigma_j).\rho_p]$ is:
\begin{equation}
    \left(
\begin{array}{ccc}
 \sin{\theta} & 0 & 0 \\
 0 & -\sin{\theta} & 0 \\
 0 & 0 & 1 \\
\end{array}
\right)
\label{CM_PS}
\end{equation}

We obtain $r_{\text{Smin}}(\rho_p)$ under optimal symmetric collective attacks(OSCA), for the pure state using Eq.(\ref{QBER_r}) and Eq.(\ref{DIrate}) in terms of negativity $N$,
\begin{align}
    r_{\text{Smin}(\rho_p)} &=  \frac{-\log64+(1-2N)\log(1-2N)+(3+2N)\log(3+2N)}{\log4}
\end{align}

For the case of collective attacks(CA), using Eq.(\ref{QBER_r}) and Eq.(\ref{DIrateC}) we obtain the  $r_{\text{Cmin}}(\rho_p)$ in terms of negativity $N$,
 \begin{align}
    r_{\text{Cmin}(\rho_p)}=&  \frac{1}{\log16}\bigg[-8\log2+(3+2N)\log(3+2N)+(3-6N)\log(1-2N)\nonumber\\
    &+(2+4N)\log(1+2N)\bigg]
 \end{align}

$\bullet$ {\bf Werner state:}

The matrix form of the Werner state (\ref{ws}) is as follows:
\begin{equation}
\rho_w = \left(
\begin{array}{cccc}
\frac{1+p}{4} & 0 & 0 & \frac{p}{2} \\
 0 & \frac{1-p}{4} & 0 & 0 \\
 0 & 0 & \frac{1-p}{4} & 0 \\
\frac{p}{2} & 0 & 0 & \frac{1+p}{4} \\
\end{array}
\right)
\end{equation}

For the Werner state, the correlation matrix are $t_{ij} = \text{Tr} [(\sigma_i \otimes \sigma_j).\rho_w]$ is:
\begin{equation}
    \left(
\begin{array}{ccc}
 p & 0 & 0 \\
 0 & -p & 0 \\
 0 & 0 & p \\
\end{array}
\right)
\label{CM_WS}
\end{equation}

We obtain $r_{\text{Smin}}(\rho_w)$ under optimal symmetric collective attacks(OSCA), for the Werner state using Eq.(\ref{QBER_r}) and Eq.(\ref{DIrate}) in terms of negativity $N$,
\begin{align}
    r_{\text{Smin}(\rho_w)} &=  \frac{\log8+(2-4N)\log(\frac{1-2N}{3})+4(1+N)\log(\frac{2(1+N)}{3})}{\log8}
\end{align}

Using Eq.(\ref{QBER_r}) and Eq.(\ref{DIrateC}) we obtain the  $r_{\text{Cmin}}(\rho_w)$ under collective attacks(CA), in terms of negativity $N$,
 \begin{align}
    r_{\text{Cmin}(\rho_w)} =&  \frac{1}{6\log2}\bigg[-12\log3+2(1-2N)\log(1-2N)+(4+4N)\log(2+2N)\nonumber\\
    &+(3-\delta)\log(3-\delta)+(3+\delta)\log(3+\delta)\bigg]
 \end{align}
 
Where, $\delta = \sqrt{-7+16N+32N^2}$.


\begin{thebibliography}{99}

\bibitem{Gisin02} N. Gisin, G. Ribordy, W. Tittel, and H. Zbinden, \emph{Quantum cryptography}, \href{https://journals.aps.org/rmp/abstract/10.1103/RevModPhys.74.145}{Rev. Mod. Phys. 74, 145 (2002).}

\bibitem{Feihu20} F. Xu, X. Ma, Q. Zhang, H.K. Lo, and J.W. Pan, \emph{Secure quantum key distribution with realistic devices},  \href{https://journals.aps.org/rmp/abstract/10.1103/RevModPhys.92.025002}{Rev. Mod. Phys. 92, 025002 (2020).}

\bibitem{Daniel17} D. J. Bernstein
 and T. Lange, \emph{Post-quantum cryptography},  \href{https://www.nature.com/articles/nature23461.pdf?origin=ppub}{Nature Vol 549, 23461 (2017).}

\bibitem{BB84} C. H. Bennett and G. Brassard, \emph{Quantum cryptography, public key distribution and coin tossing}, \href{https://www.sciencedirect.com/science/article/pii/S0304397514004241?via%3Dihub}{Theoretical Computer Science, Vol. 560 (Part 1), pp. 7-11 (2014).}

\bibitem{Bennet92} C. H. Bennett, \emph{Quantum cryptography using any two nonorthogonal states}, \href{https://journals.aps.org/prl/abstract/10.1103/PhysRevLett.68.3121}{Phys. Rev. Lett. \textbf{68}, 3121 (1992).}

\bibitem{ekert} A. K. Ekert, \emph{Quantum cryptography based on Bell's theorem
}, \href{https://journals.aps.org/prl/abstract/10.1103/PhysRevLett.67.661}{Phys. Rev. Lett. 67, 661  (1991).}



\bibitem{Wootters92} W. K. Wootters and W. H. Zurek , \emph{A single quantum cannot be cloned}, \href{https://www.nature.com/articles/299802a0}{Nature volume 299, pages802–803 (1982).}

\bibitem{Pawlowski10} M. Pawłowski, \emph{Security proof for cryptographic protocols based only on the monogamy of Bell’s inequality violations}, \href{https://journals.aps.org/pra/abstract/10.1103/PhysRevA.82.032313}{Phys. Rev. A 82, 032313 (2010).}

\bibitem{Pramanik} T. Pramanik, M. Kaplan and A. S. Majumdar, \emph{Fine-grained Einstein-Podolsky-Rosen steering inequalities}, \href{https://journals.aps.org/pra/abstract/10.1103/PhysRevA.90.050305}{Phys. Rev. A {\bf 90}, 050305(R) (2014).} 

\bibitem{Acin07} A. Acín, N. Brunner, N. Gisin, S. Massar, S. Pironio, and V. Scarani, \emph{Device-Independent Security of Quantum Cryptography against Collective Attacks}, \href{https://journals.aps.org/prl/abstract/10.1103/PhysRevLett.98.230501}{Phys. Rev. Lett. 98, 230501 (2007).} 

 \bibitem{vazirani} U. Vazirani, T. Vidick, \emph{Fully Device Independent Quantum Key Distribution},
\href{https://journals.aps.org/prl/abstract/10.1103/PhysRevLett.113.140501}{Phys. Rev. Lett. {\bf 113}, 140501 (2014)}.

\bibitem{Farkas21} M. Farkas, M. B. Juandó, K. Łukanowski, J. Kołodyński, and A. Acín, \emph{Bell Nonlocality Is Not Sufficient for the Security of Standard Device-Independent Quantum Key Distribution Protocols
}, \href{https://journals.aps.org/prl/abstract/10.1103/PhysRevLett.127.050503}{Phys. Rev. Lett. 127, 050503 (2021).} 

\bibitem{Jaskaran} J. Singh, S. Ghosh, Arvind, S. K.Goyal, \emph{Role of Bell-CHSH violation and local filtering in quantum key distribution}, \href{https://doi.org/10.1016/j.physleta.2021.127158}{Physics Letters A, Volume \bf{392}, 127158, (2021)}.







\bibitem{Jan20} J. Kołodyński, A. Máttar, P. Skrzypczyk, E. Woodhead, D. Cavalcanti, K. Banaszek, and A. Acín, \emph{Device-independent quantum key distribution with single-photon sources
}, \href{https://quantum-journal.org/papers/q-2020-04-30-260/}{Quantum 4, 260 (2020).} 



\bibitem{Metger21} T. Metger, Y. Dulek, A. Coladangelo and R. A. Friedman, \emph{Device-independent quantum key distribution from computational assumptions
}, \href{https://iopscience.iop.org/article/10.1088/1367-2630/ac304b/meta}{New J. Phys. 23 123021 (2021).} 


\bibitem{Nadlinger21} D. P. Nadlinger, P. Drmota, B. C. Nichol, G. Araneda, D. Main, R. Srinivas, D. M. Lucas, C. J. Ballance, K. Ivanov, E. Y-Z. Tan, P. Sekatski, R. L. Urbanke, R. Renner, N. Sangouard, J-D. Bancal, \emph{Experimental quantum key distribution certified by Bell's theorem
}, \href{https://www.nature.com/articles/s41586-022-04941-5}{Nature volume 607, pages 682–686 (2022).} 

\bibitem{Xu21} F. Xu, Y. Z. Zhang, Q. Zhang, J. Pan, \emph{Device-independent quantum key distribution with random post selection
}, \href{https://journals.aps.org/prl/abstract/10.1103/PhysRevLett.128.110506}{Phys. Rev. Lett. 128, 110506 (2022).} 

\bibitem{Pironio09} S. Pironio, A. Acín, N. Brunner, N. Gisin, S. Massar and V. Scarani, \emph{Device-independent quantum key distribution secure against collective attacks
}, \href{https://iopscience.iop.org/article/10.1088/1367-2630/11/4/045021}{New J. Phys. 11 045021 (2009).}

\bibitem{Ferenczi12} A. Ferenczi and N. Lütkenhaus, \emph{Symmetries in quantum key distribution and the connection between optimal attacks and optimal cloning}, \href{https://journals.aps.org/pra/abstract/10.1103/PhysRevA.85.052310}{Phys. Rev. A 85, 052310 (2012).}

\bibitem{LanZhou} L. Zhou, Y. B. Sheng, G. L. Long, \emph{Device-independent quantum secure direct communication against collective attacks}, \href{https://www.sciencedirect.com/science/article/abs/pii/S2095927319306206}{Science Bulletin, Volume \bf{65}, Issue 1, (2020).}

\bibitem{LanZhou1} L. Zhou, Y. B. Sheng, \emph{One-step device-independent quantum secure direct communication}, \href{https://doi.org/10.1007/s11433-021-1863-9}{Sci. China Phys. Mech. Astron. \bf{65}, 250311 (2022).}

\bibitem{WeiZhang} W. Zhang, D. S. Ding, Y. B. Sheng, L. Zhou, B. S. Shi, and G. C. Guo, \emph{Quantum Secure Direct Communication with Quantum Memory}, \href{https://doi.org/10.1103/PhysRevLett.118.220501}{Phys. Rev. Lett. 118, 220501 (2017).}


\bibitem{YZg} Y. Zhang, Z. Chen, S. Pirandola, X. Wang, C. Zhou, B. Chu, Y. Zhao, B. Xu, S. Yu, and H. Guo, \emph{Long-Distance Continuous-Variable Quantum Key Distribution over 202.81 km of Fiber}, \href{https://journals.aps.org/prl/abstract/10.1103/PhysRevLett.125.010502}{Phys. Rev. Lett. \bf{125}, 010502(2020)}.

\bibitem{PJLJ} J. Park, J. Lee, J. Heo, \emph{Improved statistical fluctuation analysis for twin-field quantum key distribution}, \href{https://link.springer.com/article/10.1007/s11128-021-03035-x}{Quant. Inform. Process. \bf{20}, 127 (2021).}

\bibitem{PJLJ1} Y.F. Lu, Y. Wang, M.S. Jiang, F. Liu, X.X. Zhang, W.S. Bao, \emph{Finite-key analysis of sending-or-not-sending
 twin-field quantum key distribution with intensity fluctuations}, \href{https://link.springer.com/article/10.1007/s11128-021-03070-8}{Quantum Information Processing 20, 135(2021).}
 
 \bibitem{PJLJ2} L. G. She, C. M. Zhang, \emph{Reference-frame-independent quantum key distribution with modified coherent states}, \href{https://link.springer.com/article/10.1007/s11128-022-03502-z}{Quant. Inform. Process. \bf{21}, 161 (2022).}

\bibitem{Horodecki08} K. Horodecki, M. Horodecki, P. Horodecki, D. Leung and J. Oppenheim, \emph{Quantum Key Distribution Based on Private States: Unconditional Security Over Untrusted Channels With Zero Quantum Capacity}, \href{https://ieeexplore.ieee.org/document/4529275}{IEEE Transactions on Information Theory, vol. 54, no. 6, pp. 2604-2620 (2008).}

\bibitem{Kas03} D. Kaszlikowski and Daniel K. L. Oi and M. Christandl and K. Chang and A. K. Ekert and L. C. Kwek and C. H. Oh,\emph{Quantum cryptography based on qutrit Bell inequalities},\href{https://journals.aps.org/pra/abstract/10.1103/PhysRevA.67.012310}{Phys. Rev. A 67, 012310 (2003).}

\bibitem{Ratul20} R. Banerjee, A. K. Pal, and A. Sen(De), \emph{Uniform decoherence effect on localizable entanglement in random multiqubit pure states}, \href{https://journals.aps.org/pra/abstract/10.1103/PhysRevA.101.042339}{Phys. Rev. A 101, 042339  (2020).}

\bibitem{Rivu21} R. Gupta, S. Gupta, S. Mal, and A. Sen(De), \emph{Performance of dense coding and teleportation for random states: Augmentation via preprocessing}, \href{https://journals.aps.org/pra/abstract/10.1103/PhysRevA.103.032608}{Phys. Rev. A 103, 032608  (2021).}

\bibitem{Vivien02} V. M. Kendon, K. Życzkowski, and W. J. Munro, \emph{Bounds on entanglement in qudit subsystems
}, \href{https://journals.aps.org/pra/abstract/10.1103/PhysRevA.66.062310}{Phys. Rev. A 66, 062310 (2002).}

\bibitem{Waldemar19} W. Kłobus, A. Burchardt, A. Kołodziejski, M. Pandit, T. Vértesi, K. Życzkowski, and W. Laskowski, \emph{k-uniform mixed states}, \href{https://journals.aps.org/pra/abstract/10.1103/PhysRevA.100.032112}{Phys. Rev. A 100, 032112 (2019).}

\bibitem{Gross09} D. Gross, S. T. Flammia, and J. Eisert, \emph{Most Quantum States Are Too Entangled To Be Useful As Computational Resources}, \href{https://journals.aps.org/prl/abstract/10.1103/PhysRevLett.102.190501}{Phys. Rev. Lett. 102, 190501 (2009).}

\bibitem{Soorya19} S. Rethinasamy, S. Roy, T. Chanda, A. Sen(De), and U. Sen, \emph{Universality in distribution of monogamy scores for random multiqubit pure states}, \href{https://journals.aps.org/pra/abstract/10.1103/PhysRevA.99.042302}{Phys. Rev. A 99, 042302 (2019).}



\bibitem{Benjamin96} B. Schumacher and M. A. Nielsen, \emph{Quantum data processing and error correction}, \href{https://journals.aps.org/pra/abstract/10.1103/PhysRevA.54.2629}{Phys. Rev. A 54, 2629 (1996).}

\bibitem{Rivu20} R. Gupta, S. Gupta, S. Mal and A. Sen(De), \emph{Constructive Feedback of Non-Markovianity on Resources in Random Quantum States}, \href{https://journals.aps.org/pra/abstract/10.1103/PhysRevA.105.012424}{Phys. Rev. A 105,012424 (2022).}

\bibitem{Schwonnek21} R. Schwonnek, K. T. Goh, I. W. Primaatmaja, E. Y. Z. Tan, R. Wolf, V. Scarani and C. C. W. Lim, \emph{Device-independent quantum key distribution with
random key basis}, \href{https://www.nature.com/articles/s41467-021-23147-3}{Nature Communications volume 12, Article number: 2880 (2021).}






\bibitem{YMX01} Y. M. Xie, B. H. Li, Y. S. Lu, X. Y. Cao, W. B. Liu, H. L. Yin, and Z. B. Chen, \emph{Overcoming the rate–distance limit of device-independent quantum key distribution}, \href{https://opg.optica.org/ol/abstract.cfm?uri=ol-46-7-1632}{Opt. Lett. 46,1632 (2021).}

\bibitem{WZL01} W. Z. Liu, Y. Z. Zhang, Y. Z. Zhen, M. H. Li, Y. Liu, J. Fan, F. Xu, Q. Zhang, and J. W. Pan, \emph{Toward a Photonic Demonstration of Device-Independent Quantum Key Distribution}, \href{https://journals.aps.org/prl/abstract/10.1103/PhysRevLett.129.050502}{Phys. Rev. Lett. 129, 050502 (2022).}

\bibitem{Wiseman07} H. M. Wiseman, S. J. Jones, and A. C. Doherty, \emph{Steering, Entanglement, Nonlocality, and the Einstein-Podolsky-Rosen Paradox}, \href{https://journals.aps.org/prl/abstract/10.1103/PhysRevLett.98.140402}{Phys. Rev. Lett. 98, 140402 (2007).}
















\bibitem{binghongli} B. H. Li, Y. M. Xie, Z. Li, C. X. Weng, C. L. Li,
H. L. Yin, and Z. B. Chen, \emph{Long-distance twin-field quantum key distribution
with entangled sources},\href{https://opg.optica.org/ol/abstract.cfm?uri=ol-46-22-5529}{Opt. Lett. {\bf 46}, 5529 (2021)}.


\bibitem{yaofu} Y. Fu, H. L. Yin, T. Y. Chen, and Z. B. Chen, \emph{Long-Distance Measurement-Device-Independent Multiparty Quantum Communication},\href{https://journals.aps.org/prl/abstract/10.1103/PhysRevLett.114.090501}{Phys. Rev. Lett. {\bf 114}, 090501 (2015)}.

\bibitem{jiegu} J. Gu, Y. M. Xie, W. B. Liu, Y. Fu, H. L. Yin, and Z. B. Chen, \emph{Secure quantum secret sharing without signal disturbance monitoring },\href{https://opg.optica.org/oe/fulltext.cfm?uri=oe-29-20-32244&id=459775}{Opt. Express {\bf 29}, 32244 (2021)}.

\bibitem{zhaoli} Z. Li, X. Y. Cao, C. L. Li, C. X. Weng, J. Gu, H. L. Yin and Z. B. Chen, \emph{Finite-key analysis for quantum conference key agreement with asymmetric channels},\href{https://iopscience.iop.org/article/10.1088/2058-9565/ac1e00/meta}{Quant. Sci. Technol. {\bf 6}, 045019 (2021)}.


\bibitem{AM1} A. Maitra, G. Paul, and S. Roy, \emph{Device-independent quantum private query}, \href{https://journals.aps.org/pra/abstract/10.1103/PhysRevA.95.042344}{Phys. Rev. A 95, 042344 (2017).}








\end{thebibliography}
\end{document}